\documentclass[usenatbib,useAMS]{mn2e}
\usepackage{graphicx}

\def\gapprox{\lower.4ex\hbox{$\;\buildrel >\over{\scriptstyle\sim}\;$}}
\def\lapprox{\lower.4ex\hbox{$\;\buildrel <\over{\scriptstyle\sim}\;$}}

\def\arctanh{{\rm arctanh}}

\def\bI{\mbox{\boldmath $I$}}
\def\bj{\mbox{\boldmath $j$}}
\def\bkappa{\mbox{\boldmath $\kappa$}}
\def\bJ{\mbox{\boldmath $J$}}

\def\bu{\mbox{\boldmath $u$}}
\def\bv{\mbox{\boldmath $v$}}

\def\br{\mbox{\boldmath $r$}}

\def\bq{\mbox{\boldmath $q$}}
\def\bB{\mbox{\boldmath $B$}}
\def\bE{\mbox{\boldmath $E$}}
\def\bA{\mbox{\boldmath $A$}}

\def\bOmega{\mbox{\boldmath $\Omega$}}
\def\bnabla{\mbox{\boldmath $\nabla$}}
\def\bbeta{\mbox{\boldmath $\beta$}}

\title[Oscillatory polar gaps]{Oscillating pulsar polar gaps}

\author[Q. Luo and D. Melrose]
      {Qinghuan Luo and Don Melrose\\
        School of Physics, The University of Sydney, NSW 2006, Australia\\
}
\date{
          --- Received in original form September, 2007
        }
\pubyear{2008}

\begin{document}

\maketitle

\begin{abstract}
An analytical model for oscillating pair creation above the pulsar polar cap
is presented in which the parallel electric field is treated as a large amplitude,
superluminal, electrostatic wave. An exact formalism for such wave is derived in 
one-dimension and applied to both the low-density regime in which the pair plasma density 
is much lower than the corotating charge density and the high-density regime in which
the pair plasma density is much higher than the corotating charge density. In the low-density regime,
which is relevant during the phase leading to a pair cascade, a parallel electric field 
develops resulting in rapid acceleration of particles. The rapid acceleration leads 
to bursts of pair production and the system switches to the oscillatory phase, 
corresponding to the high density regime, in which pairs oscillate with net drift 
motion in the direction of wave propagation.  Oscillating pairs lead to a 
current that oscillates with large amplitude about the Goldreich-Julian current.
The drift motion can be highly relativistic if the phase speed of large amplitude waves
is moderately higher than the speed of light. Thus, the model predicts 
a relativistic outflow of pairs, a feature that is required for avoiding overheating
of the pulsar polar cap and is also needed for the pulsar wind.
\end{abstract}


\section{Introduction}

One of the central problems in pulsar electrodynamics is the production of 
the relativistic electron-positron pair plasma in which coherent radio emission is 
thought to be produced \citep{s71}. It is widely believed that particles are
accelerated to ultra high energy along open field lines, leading to a 
cascade producing the pair plasma. The pair cascades should 
produce detectable X-rays and gamma-rays 
in the case of fast rotating, young pulsars and millisecond pulsars \citep{t01}.
Different acceleration locations in the pulsar magnetosphere
have been postulated, with a common feature that the acceleration results from
a large scale electric field on open field lines that extend beyond
the light cylinder. Acceleration regions
in a pulsar magnetosphere are referred to as `gaps'. 
A widely-discussed acceleration region is near the polar cap, for
which there is a class of acceleration models called polar gap models \citep{s71,as79,hm98}. 
Particle acceleration near the polar cap is of particular relevance for pulsar radio emission since
the radio observations suggest that for many pulsars the emission 
originates deep inside the pulsar magnetospheres \citep{betal91,ew01}. Other acceleration sites
include regions in the outer magnetosphere near the null surface, referred to as outer gaps 
\citep{cetal86,r96,h06} and a variant, referred to as slot gaps,
characterized by a long, thin region along the last open field lines \citep{a83,hm05}. 
Here we concentrate on the polar cap region.

The conventional polar gap models were developed on the basis that
the system can settle into a steady state, such that all 
physical quantities in the pulsar's corotating frame can be regarded as
time-independent \citep{as79,hm98}. This time-independent assumption is not  
realistic in practice because it ignores inductive electric fields. There is a strong 
argument that a time-dependent inductive field plays a central role in pulsar electrodynamics. 
A global current must be present in the system to form a current closure \citep{sw73,m75,cr76,s91} and
the steady-state assumption requires that the current density, $\bJ$,  balance $\bnabla\times\bB/\mu_0$. 
Moreover, the global electrodynamics requires that the current be determined globally, rather than
by local processes near the polar cap \citep{s91,t06}. The global and local requirements on the current density are generally incompatible without an additional source of charge and current, and
any resulting mismatch between $\bJ$ and $\bnabla\times\bB/\mu_0$ implies a time-dependent electric field. Such a mismatch and associated inductive electric field seem unavoidable.

Although the idea that a cascade above the polar cap is intrinsically time-dependent 
was suggested much earlier by \citet{s71} and adopted in \citet{rs75}'s 
spark model, there are no quantitative models that take into account the time dependence. 
\citet{letal05} recently discussed an oscillatory gap model to illustrate
the time dependent nature of the polar gap in the one-dimensional approximation. 
In this model, induction currents due to temporal changes in the system
are included and particle acceleration is shown to settle into an oscillatory state, 
similar to a large amplitude wave. The model has several limitations.
First, escape of particles is included only implicitly, and escape needs to be made explicit to be consistent with observations of pulsar winds. Second,
the oscillatory model is based on numerical integration of the relevant fluid equations
together with Maxwell's equations in one dimension and due to 
the limit of numerical calculation the oscillations could be followed only for a limited
number of periods. Third, the assumption that the oscillations are purely temporal
is unrealistic, and needs to be generalized to allow outward propagating waves.

In this paper we adopt a different approach, treating the oscillations as
a large-amplitude electrostatic wave (LAEW) in a cold pair plasma. Although LAEW in a
cold electron gas was discussed in \citet{aetal75}, to our knowledge there has been
no discussion of the case of a strongly magnetized, electron-positron pair plasma. 
Relativistic motion of a single particle in a LAEW was discussed and applied to
pulsar emission by \citet{r92a,r92b}. However, in the single-particle treatment, the wave was 
assumed to pre-exist and feedback of the particle's motion on the wave field was not taken into account. 
Here we treat the electrons and positrons as cold fluids and determine particle acceleration by solving simultaneously 
both fluid equations and Maxwell's equations without making an a priori 
assumption of time-independence. We include pair creation in our equations, but neglect it in deriving analytic solutions.

In Sec~\ref{time-dependent} we outline the fluid formalism for time-dependent 
electrodynamics that includes inductive electric fields. Analytical solutions
for LAEW are described in Sec~\ref{LAEW}. Inclusion of pair production in LAEW 
is discussed Sec~\ref{pair-production} and the low-density limit is discussed 
in Sec~\ref{low-density}.

\section{Time-dependent formalism}
\label{time-dependent}

\subsection{Fluid equations}

We consider a two-component cold fluid consisting of electrons and
positrons; the fluid number density and velocity are denoted by $N_\pm$ and $\bv_\pm$, where the 
subscripts $\pm$ correspond respectively to the positron and electron components.
In the observer's inertial frame, a pulsar rotates with an angular velocity
$\Omega=2\pi/P$, where $P$ is the pulsar period. Provided that sufficient charge density is available, the 
corotating electric field is set up so that charged particles corotate with the star. 
This electric field can be eliminated by choosing a frame corotating with the star. 
Well inside the light cylinder, the corotating frame and observer's inertial frame are connected
by a local Galilean transformation with velocity $\bv_R=\bOmega\times\br$, where $\br$ is the radial 
vector directed from the star's center to a point of interest. 
One has $|v_R/c|\sim r/R_{LC}$, with $R_{LC}=c/\Omega$ the light-cylinder radius. 
Thus, the effect on the rotation of the fluids can be
neglected if the region concerned is close to the polar cap where $r/R_{LC}\ll1$.

The relevant fluid equations, the continuity equation and equation of motion, can
be written down as
\begin{equation}
{\partial N_\pm\over\partial t}+\bnabla\cdot(N_\pm\bv_\pm)={\textstyle{1\over2}}Q,
\label{eq:ContinuityEq}
\end{equation}
\begin{equation}
\biggl({\partial\over\partial t}+\bv_\pm\cdot\bnabla\biggr)
\bu_\pm=\pm{e\bE\over m_ec}+{\bq_\pm\over m_ec^2}-{Q\over 2N_\pm}\bu_\pm,
\label{eq:EqMotion}
\end{equation}
where $\bq_\pm$ is radiation drag, $Q$ is a source function due to pair production and
$\bu_\pm=\gamma_\pm\bv_\pm/c$ is the particle's dimensionless momentum.
The Lorentz force is absent because all the particles are assumed to be 
in the ground Landau state. The current and charge densities are
\begin{equation}
\bJ=e\sum_{s=\pm}s\bv_sN_s,\quad \rho=e\sum_{s=\pm}sN_s,
\end{equation}
with $s=\pm$. The fluid equations are supplemented by Maxwell's equations, written
in the corotating frame. The two relevant Maxwell equations 
involve the charge density and current density \citep{fetal77}
\begin{eqnarray}
\bnabla\cdot\bE&=&{1\over\varepsilon_0}(\rho-\rho_{_{GJ}}),\label{eq:divE}\\
\bnabla\times\bB&=&\mu_0(\bJ-\bJ_R)+{1\over c^2}{\partial\bE\over\partial t},
\label{eq:curlB}
\end{eqnarray}
where $\rho_{_{GJ}}=\varepsilon_0[ -2\bOmega\cdot\bB+(\bOmega\times\br)\cdot(\bnabla\times\bB)]$
is the Goldreich-Julian (GJ) density. Here we ignore the general relativistic effects such as 
frame dragging (cf. Sec 5.2). In (\ref{eq:curlB}) $\bJ_R$ is a vectorial sum of all
the remaining terms that are small for $r/R_{LC}\ll1$; the full expression for $\bJ_R$ is 
given by (A5) in \citet{fetal77} and it is neglected here.
Equation (\ref{eq:divE}) describes a noncorotating electric field arising 
from deviation of charge density from the GJ density.  In steady-state models, equation  
(\ref{eq:curlB}) is (implicitly) assumed to be satisfied trivially.  However, it plays 
a central role here in determining the inductive field arising from a current mismatch. 

The inclusion of the inductive field distinguishes the 
model considered here from steady-state polar-gap models in which only 
Poisson's equation  (i.e., Eq \ref{eq:divE}) is relevant and the parallel electric field
is treated as static \citep{as79,hm98}. The static assumption is incompatible, in general, with
the constraint imposed by a global current. It has long been recognized 
that circulation of a global current plays a critical role in dissipation of 
rotational energy of pulsars \citep{sw73,m75,cr76}. Such current, denoted by 
$\bJ_0$, should be determined globally. The simplest case is where the global current is assumed to be
a constant. Assuming $\bJ_{0\parallel}=(\bnabla\times\bB)_\parallel$, where
$\parallel$ represents projection along the magnetic field,  the steady-state assumption 
implies that the local current exactly matches the global current, $\bJ_\parallel=\bJ_{0\parallel}$
(when $\bJ_R$ is ignored), everywhere.
This assumption is not realistic, for example, due to pair creation changing the current density 
locally \citep{letal05}. In our oscillatory model, the parallel electric field is 
treated predominantly as an inductive field due to oscillation of the current about the 
global constant direct current. Relevant solutions
are oscillatory and should be treated as large amplitude waves. 

\subsection{Large amplitude waves}

We outline a general approach for deriving a time-dependent solution in which 
oscillations are considered as a large amplitude wave propagating in the direction 
$\bkappa$ at a constant phase speed $\beta_V$ (in units of $c$). We assume that oscillatory 
quantities are functions of
\begin{equation}
\chi=\omega_{_{GJ}}\left(\beta_Vt-{\bkappa\cdot\br\over c}\right),
\end{equation}
where $\omega_{_{GJ}}=(e^2N_{GJ}/\varepsilon_0m_e)^{1/2}$ 
is the plasma frequency at the GJ number density, 
$N_{GJ}=2\varepsilon_0\Omega B_0/e$, and $B_0$ is the surface magnetic field.
We normalize the electric field as $\tilde{E}_\parallel=eE_\parallel/m_ec\omega_{_{GJ}}$, density
as $\tilde{N}_\pm=N_\pm/N_{GJ}$, the charge density as $\eta=\rho/ecN_{GJ}$ and 
the current density as $\bj=\bJ/ecN_{GJ}$. Using $\partial/\partial t=\omega_{GJ}\beta_Vd/d\chi$ and 
$\bnabla\equiv\partial/\partial \br=-(\omega_{GJ}\bkappa/c) d/d\chi$, one may write
(\ref{eq:ContinuityEq}),  (\ref{eq:EqMotion}) and (\ref{eq:curlB}) 
in the dimensionless forms
\begin{equation}
{d\over d\chi}\biggl[(\beta_V-\bkappa\cdot\bbeta_\pm)\tilde{N}_\pm\biggr]
={\textstyle{1\over2}}\tilde{Q},
\label{eq:ContinuityEq2}
\end{equation}
\begin{equation}
(\beta_V-\bkappa\cdot\bbeta_\pm){d\bu_\pm\over d\chi}=\pm\tilde{\bE}
+\tilde{\bq}_\pm-{\tilde{Q}\over 2\tilde{N}_\pm}\bu_\pm,
\label{eq:EqMotion2}
\end{equation}
\begin{equation}
\Bigl[\bigr(1-\beta^2_V\bigl)\bI-\bkappa\bkappa\Bigl]\cdot{d\tilde{\bE}\over d\chi}
=\beta_V\bigl(\bj-\bj_0\bigr),
\label{eq:waveE2}
\end{equation}
where $\bbeta_\pm=\bv_\pm/c$, $\tilde{\bq}_\pm=\bq_\pm/(m_ec^2\omega_{_{GJ}})$,
$\tilde{Q}=Q/(N_{GJ}\omega_{_{GJ}})$, $(\bI)_{ij}=\delta_{ij}$ and 
$\bj_0=\bnabla\times\bB/(eN_{GJ}c)$ which is assumed to be a constant vector.
Poisson's equation (4) and the induction equation (5) may be replaced by the equation of charge
continuity and the induction equation. With the charge and current densities 
functions only of $\chi$, the equation of charge continuity gives
\begin{equation}
{d\over d\chi}(\beta_V\eta-{\bkappa}\cdot\bj)=0.
\label{eq:cce}
\end{equation}
Note that in our model both Poisson's equation and the induction equation contribute. The steady state
models correspond to the limit $\beta_V\to0$, and hence $\bkappa\cdot\bj=$ const., when the induction equation
does not contribute. It is only in the opposite limit, $\beta_V\to\infty$, where the oscillations are
purely temporal, that Poisson's equation makes no contribution (cf. Sec 2.3).
As we are primarily interested in oscillatiory 
solutions, in deriving these equations we ignore inhomogeneity, notably in the
magnetic field. This neglect is justified provided that the
spatial scale for the inhomogeneity  is much larger 
than the oscillation length. The scale of the inhomogeneity is of order the radius of curvature,
which at a height $r-R$ with $R=10^4\,\rm m$ the star's radius, is larger than but of order $(rR_{LC})^{1/2}$. One has
$\sim (rR_{LC})^{1/2}\approx 2.2\times10^5(r/R)^{1/2}P_{0.1}^{1/2}\,\rm m$, where
$P_{0.1}=P/0.1\,\rm s$. 
For plausible parameters, this is much larger than the oscillation length, which is of order 
\begin{eqnarray}
\lambda&=&{c\beta_V\over\omega_{GJ}}\left({
\gamma\over\tilde{N}}\right)^{1/2}\nonumber\\
&\approx&
0.9\beta_VB^{1/2}_8P^{-1/2}_{0.1}\left({\gamma\over10^6}\right)^{1/2}\left({\tilde{N}\over
10^2}\right)^{-1/2}\,{\rm m}, 
\label{eq:lambda}
\end{eqnarray}
with $B_8=B_0/10^8\,\rm T$, $\tilde{N}={\rm max}\{\tilde{N}_+,\tilde{N}_-\}$ and where
we asssume an oscillation frequency $\sim\omega_p/\gamma^{1/2}$
(see Eq \ref{eq:freq}). For the numerical example in (\ref{eq:lambda}), the condition is satisfied 
provided that $\beta_V<2.4\times10^5$. We are only concerned with the open field line region
which can be regarded as a flux tube with a conducting surface, defined by the last
closed field lines, that separates the region from the closed field line region.
Eq (\ref{eq:ContinuityEq2})--(\ref{eq:waveE2}) are then valid only
when the wavelength is much shorter than the transverse size 
$\sim (r/R_{LC})^{1/2}r\geq 258 P_{0.1}^{-1/2}\,\rm m$.
Such short wavelength approximation implies that the limit $\beta_V\to\infty$ is not 
applicable for pulsars. Nonetheless, such limit is also 
discussed here as it simplies the formalism from which the basic properties of
LAEWs can be derived and compared to a more general case where $\beta_V$ is finite.

Integration of Eq (\ref{eq:ContinuityEq2}) yields an exact form for the density
\begin{equation}
\tilde{N}_\pm={(\beta_V-\bkappa\cdot\bbeta_{\pm0})n_\pm+F_Q\over \beta_V-\bkappa\cdot\bbeta_\pm},
\quad F_Q={\textstyle{1\over2}}\!\!\int^\chi_0 \tilde{Q}(\chi')\,d\chi',
\label{eq:Npm}
\end{equation}
where $n_\pm=\tilde{N}_{\pm}(0)$ is the initial density at $\chi=0$, 
$\bbeta_{\pm0}$ are the initial velocities  (in units of $c$), and 
$F_Q$ is a cumulative flux arising from pair creation. A wave is classified as
superluminal if $\beta_V>1$, luminal if $\beta_V=1$ and subluninal if 
$\beta_V<1$. We do not consider the subluminal case here. 
For superluminal and luminal waves, $\tilde{N}_\pm$ is always
positive. The number density remains approximately constant in 
a superluminal wave in the limit $\beta_V\to\infty$.

\subsection{Current-charge invariant}

The charge continuity equation (\ref{eq:cce}) implies an invariant 
$\bkappa\cdot\bj-\beta_V\eta={\rm const}$.  Denoting the dimensionless current density by 
$j_\parallel=\bkappa\cdot\bj$, this conservation law implies
\begin{equation}
j_\parallel(\chi)-\beta_V\eta(\chi)=j_\parallel(0)-\beta_V\eta(0),
\label{eq:j-eta1}
\end{equation}
where $\chi=0$ corresponds to the initial conditions, with
$j_\parallel(0)=\beta_{+0}n_+-\beta_{-0}n_-$, $\eta(0)=n_+-n_-$.
The steady state condition corresponds to the special
limit $\beta_V\to0$, and in this limit the induction equation
is satisfied trivially with
$j_\parallel=j_{0\parallel}$.
The constant current $\bj_0$ is interpreted as the global direct current, assumed
to be determined by global conditions and to be a free parameter in the theory.
In steady state models, only Poisson's equation is relevant and the constant 
current $j_{0\parallel}$ does not appear explicitly. 
For $\beta_V\ne0$, it is convenient to
write $\eta_{_{GJ}}=\rho_{_{GJ}}/eN_{GJ}$, for the sign of the GJ charge density,
and to write the right hand side of Eq (\ref{eq:j-eta1}) in the form 
\begin{equation}
j_\parallel(0)-\beta_V\eta(0)=
j_{0\parallel}-\beta_V\eta_{_{GJ}}.
\label{eq:j-eta2}
\end{equation}
The opposite limit $\beta_V\to\infty$ corresponds to purely temporal
oscillations. In this case (\ref{eq:j-eta2}) requires that the charge
density equal the GJ density, so that Poisson's equation is satisfied
trivially. We are interested in the general case $0<\beta_V<\infty$.

Assuming a strong magnetic field approximation, a plausible simplifying assumption is that $\bkappa$ 
is directed along the magnetic field. The problem then becomes one dimensional, 
involving only projections of the relevant equations along the magnetic field. 
Of particular significance is $j_{0\parallel}$, which would be identically 
zero if the magnetic field were dipolar. The global requirement for $j_{0\parallel}\ne0$ implies a 
nonzero azimuthal magnetic field $B_\phi\sim (r/R_{LC})j_{0\parallel}B\ll B$ (see further 
discussion in Sec 3.4). Although $j_{0\parallel}$ cannot be determined locally, it 
is plausible to assume that it has the same sign as $\eta_{GJ}$, i.e.,
$j_{0\parallel}<0$ for $\bOmega\cdot\bB>0$ and $j_{0\parallel}>0$ for $\bOmega\cdot\bB<0$.

One may use (\ref{eq:j-eta1}) and (\ref{eq:j-eta2}) to express $\beta_V$ in terms of the initial 
density and velocity:
\begin{equation}
\beta_V={j_{0\parallel}-\beta_{+0}n_++\beta_{-0}n_-\over
\eta_{_{GJ}}-n_++n_-},
\label{eq:j0}
\end{equation}
provided that $\eta(0)=n_+-n_-\neq\eta_{GJ}$. The following three cases are of interest: 
the initial charge density matches the GJ density $\delta\eta=\eta_{GJ}-\eta(0)=0$,
the initial charge density has a small deviation from the GJ density 
$|\delta\eta|\ll1$, and pair density is much lower than the GJ density $n_\pm\ll1$.
In the first case, one must have $j_\parallel(0)=j_{0\parallel}$ and thus,
the phase speed is not constrained by (\ref{eq:j-eta2}).
Generally, the second applies to polar cap regions where both frame-dragging and 
field line curvature may cause a small deviation of a charge density from the corotation density
\citep{as79,s97,hm98}. Eq (\ref{eq:j0}) implies $|\beta_V|\gg1$ for $|\delta\eta|\ll|1-\beta_0|$, 
$j_{0\parallel}=\eta_{GJ}$ and $\beta_0=\beta_{+0}\approx\beta_{-0}$. Therefore, the 
LAEW considered here should be superluminal. In the third case, one has $\beta_V\approx j_{0\parallel}/\eta_{GJ}$;
the wave is superluminal for $j_{0\parallel}>\eta_{GJ}$ and subluminal for $j_{0\parallel}<\eta_{GJ}$.   

\section{Large amplitude, electrostatic waves}
\label{LAEW}

We consider large amplitude, electrostatic waves in a high density regime where the 
density of pairs is much higher than the GJ density, $n_\pm\gg1$, and
pair creation is absent. This regime is especially applicable when the system 
undergoes a brief burst of pair production leading to a pair plasma with $n_\pm\gg1$ 
and sets up oscillations. 

\begin{figure}
\includegraphics[width=7cm]{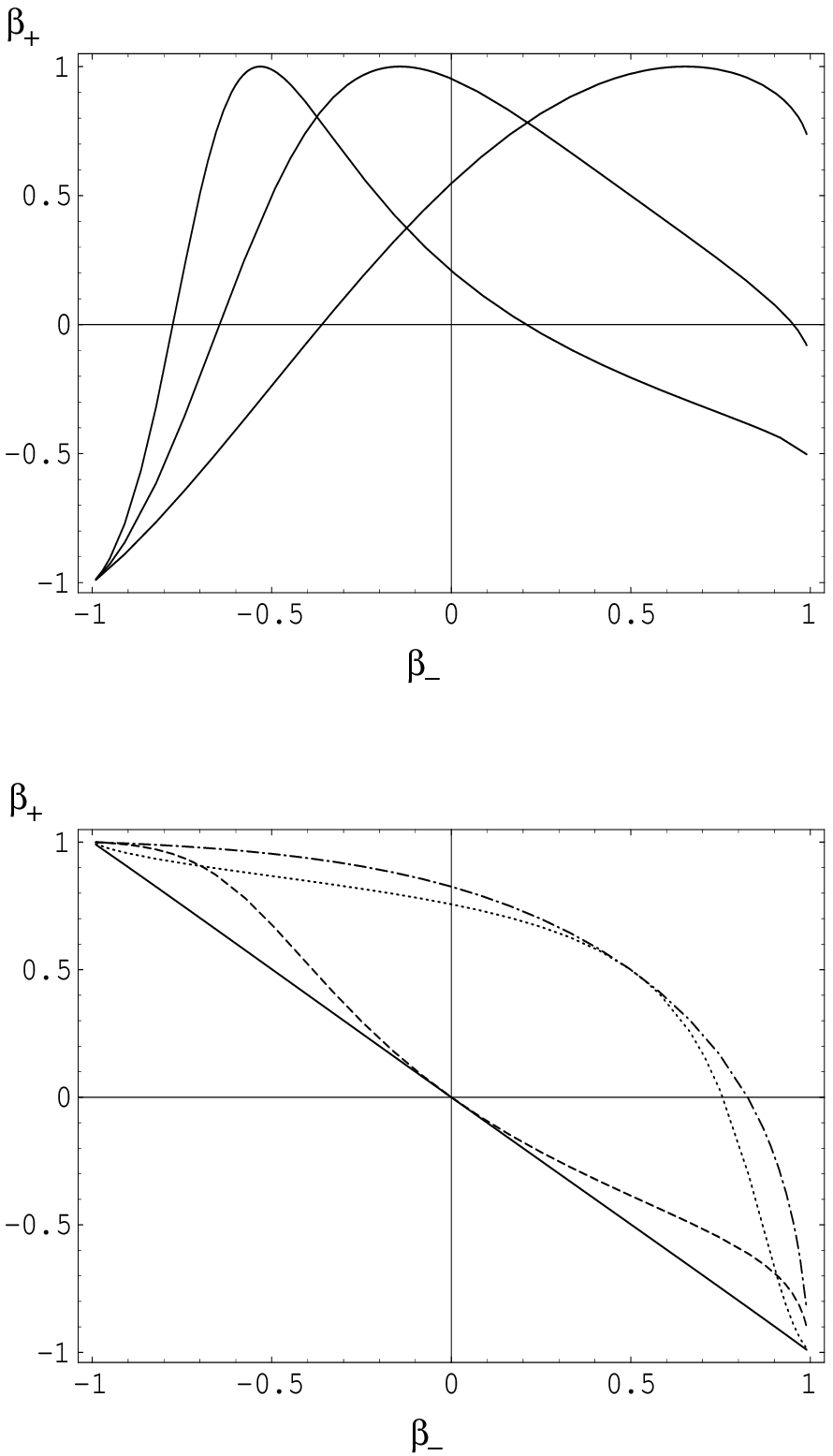}
\caption{Plot of $\beta_+$ as a function of $\beta_-$. Upper: $\beta_V=1$. The curves from 
left to right correspond to $\beta_{\pm0}=0$, $0.5$ and $0.9$, respectively. 
In each cases, the physical range corresponds to a range from where 
a maxima occurs to the rightmost.  Lower: $\beta_V=1.5$ and $\beta_{\pm0}=0$ (dashed),
$\beta_V=1.5$ and $\beta_{\pm0}=0.5$ (dash-dotted), $\beta_V=100$ and $\beta_{\pm0}=0$
(solid), $\beta_V=100$ and $\beta_{\pm0}=0.5$ (dotted).
}
\label{fig:betap}
\end{figure}

\begin{figure}
\includegraphics[width=7cm]{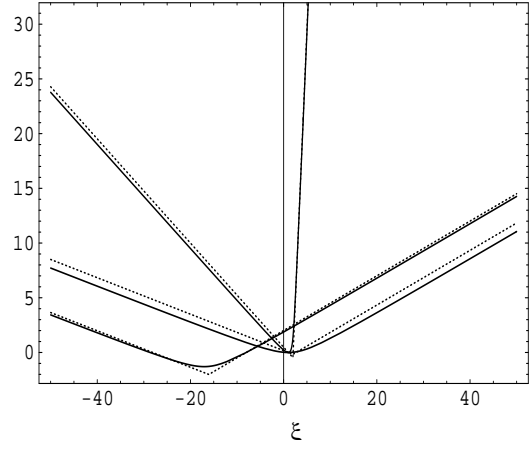}
\caption{Plot of $g(\xi)$. The three solid V-shaped curves from left to right correspond
respectively to $\beta_V=5$ with $\beta_{\pm0}=0.9$, $\beta_V=5$ with $\beta_{\pm0}=0$, and
$\beta_V=1.1$ with $\beta_{\pm0}=0$. The minima where the two lines meet are located
at $\xi_1\approx-16$ for the leftmost plot and $\xi=\xi_1=2$ for the other two. 
The dotted lines are obtained using the approximation (\ref{eq:g2}) which consists of two straight lines 
with gradients $1/(\beta_V-1)$ and $-1/(\beta_V+1)$, respectively.}
\label{fig:g}
\end{figure}

To concentrate on the basic physics of such large amplitude waves
we ignore the radiation drag, $\tilde{q}_\pm=0$. 
Eqs (\ref{eq:waveE2}) and (\ref{eq:EqMotion2}) reduce to the following simple forms:
\begin{eqnarray}
{d\tilde{E}_\parallel\over d\chi}={j_{0\parallel}\over\beta_V}
-{1\over\beta_V}\sum_{s=\pm}s\beta_s{(\beta_V-\beta_{s0})n_s
\over\beta_V-\beta_s}.
\label{eq:waveE3}
\end{eqnarray}
\begin{equation} 
{d\over d\chi}\,{1-\beta_V\beta_\pm\over(1-\beta^2_\pm)^{1/2}}=
\mp\tilde{E}_\parallel.
\label{eq:EqMotion3}
\end{equation}
Integration of (\ref{eq:EqMotion3}) leads to the following invariant:
\begin{eqnarray}
\sum_{s=\pm}{1-\beta_V\beta_s\over(1-\beta^2_s)^{1/2}}
=\sum_{s=\pm}{1-\beta_V\beta_{s0}\over(1-\beta^2_{s0})^{1/2}}\equiv\xi_1.
\label{eq:betapm}
\end{eqnarray}
One needs to consider only one component of the fluid, say the electron ($-$) 
and the solution for the other component can be derived using (\ref{eq:betapm}). 
Plots of $\beta_+$ as a function of $\beta_-$ are shown in figure 
\ref{fig:betap} for luminal and superluminal waves. Using the 
notation 
\begin{equation}
\xi\equiv{1-\beta_V\beta_-\over(1-\beta^2_-)^{1/2}},
\end{equation}
the velocity can be expressed in terms of $\xi$:
\begin{eqnarray}
\beta_-(\xi)&=&{\beta_V-\xi\left(\beta^2_V+\xi^2-1\right)^{1/2}\over
\beta^2_V+\xi^2},
\label{eq:betam}\\
\beta_+(\xi)&=&\beta_-(\xi_1-\xi).
\label{eq:betap}
\end{eqnarray}
Notice that the two velocities (\ref{eq:betam}) and (\ref{eq:betap}) are related by a 
transform $\xi\to \xi_1-\xi$.  Exact solutions to (\ref{eq:waveE3}) and (\ref{eq:EqMotion3}) 
are 
\begin{eqnarray}
-\int\,{d\xi\over\Phi^{1/2}(\xi)}=\chi,
\label{eq:EqMotion4}
\end{eqnarray}
\begin{eqnarray}
\tilde{E}_\parallel=\pm\Phi^{1/2},
\label{eq:waveE4}
\end{eqnarray}
with
\begin{eqnarray}
\Phi(\xi)&=&\tilde{E}^2_0+{2\over\beta_V}
\biggl[(\xi-\xi_0)j_0-(\beta_V-\beta_{+0})g(\xi)n_+\nonumber\\
&&-(\beta_V-\beta_{-0})\Bigl(\gamma_-(\xi)-\gamma_{-0}\Bigr)n_{-}\biggr],
\label{eq:Phi}
\end{eqnarray}
\begin{eqnarray} 
g&=&\int^\xi_{\xi_0}{\beta_+(\xi')\over\beta_V-\beta_+(\xi')}\,d\xi'\nonumber\\
&=&-{1\over\beta^2_V-1}\Biggl\{\xi_0-\xi-\beta_V\biggl[
\Bigl(\beta^2_V+(\xi_1-\xi)^2-1\Bigr)^{1/2}\nonumber\\
&&-
\Bigl(\beta^2_V+(\xi_1-\xi_0)^2-1\Bigr)^{1/2}\biggr]\Biggr\},
\label{eq:g}
\end{eqnarray}
where $\xi_0=\xi(0)$ and $\gamma_{-0}=\gamma_-(\xi_0)$ is the initial 
Lorentz factor of electrons. The initial electric field is 
$\tilde{E}_0=\pm\Phi^{1/2}(\xi_1)=d\xi/d\chi$ at $\chi=0$,
where the sign is determined by the sign of $d\xi/d\chi$ at $\chi=0$. 
Although we are interested in superluminal waves, the calculation up to this stage 
applies to any $\beta_V$ including the special case $\beta_V=1$. 
For $\beta_V=1$, (\ref{eq:g}) simplifies to $g=[(\xi_1-\xi)^{-1}-(\xi_1-\xi_0)^{-1}+\xi_0-\xi]/2$.
The function $g(\xi)$ is shown in figure~\ref{fig:g} for superluminal waves $\beta_V>1$.

\begin{figure} \includegraphics[width=7cm]{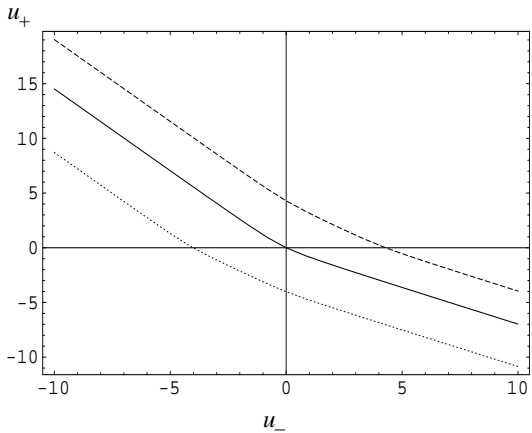}
\caption{Plot of $u_+$ as a function of $u_-$ for $\beta_{0\pm}=0$ (solid),
$\beta_{0\pm}=0.9$ (dashed) and $\beta_{0\pm}=-0.9$ (dotted). Because
of $\xi\neq0$, motions of electrons and positrons are not strictly 
symmetric about $u_+\to -u_-$.}
\label{fig:up}
\end{figure}

\subsection{Condtions for an oscillatory solution}

The condition for existence of an oscillatory solution for (\ref{eq:waveE3}) and (\ref{eq:EqMotion3}) 
is $\Phi(\xi)\geq0$. For $n_\pm\gg 1$, (\ref{eq:Phi}) simplifies to 
\begin{equation}
\Phi\approx \tilde{E}^2_0-(2/\beta_V)(\beta_V-\beta_{-0})(g+\gamma_--\gamma_0)n_-
\geq0, 
\label{eq:Phi2}
\end{equation}
where (\ref{eq:j0}) is used to eliminate $n_+$.
An appropriate approximation to $g(\xi)$ can be derived as follows.
In each oscillation a particle's velocity is highly relativistic except for a 
short phase when it is briefly nonrelativistic and changes sign.
In this case, (\ref{eq:g}) can be approximated by two straight lines that cross
at $\xi\approx\xi_1$. For $\beta_{\pm0}=0$ (solid and dashed) the minimum is located at
$\xi=\xi_1=2$. Specifically, for $(\xi_1-\xi_0)^2\gg\beta^2_V-1$ and $(\xi_1-\xi)^2\gg\beta^2_V-1$,
one has the following approximation
\begin{equation}
g(\xi)\approx\left\{
\begin{array}{ll}
\displaystyle{{\xi_0-\xi_1\over1+h(\xi_1-\xi_0)\beta_V}+{\xi-\xi_1\over\beta_V-1}},&\xi\geq\xi_1,\\
\displaystyle{{\xi_0-\xi_1\over1+h(\xi_1-\xi_0)\beta_V}-{\xi-\xi_1\over\beta_V+1}},&\xi<\xi_1,
\end{array}
\right.
\label{eq:g2}
\end{equation}
with $h(x)=1$ for $x\geq0$ and $h(x)=-1$ for $x<0$.  In the large 
$\beta_V$ limit $g(\xi)$ is symmetric about the vertical axis 
at $\xi=\xi_1$. As the phase velocity approaches the luminal limit $\beta_V\to1$ the 
right-hand side of the curve steepens and approached the vertical axis.
In figure \ref{fig:g}, (\ref{eq:g2}) is shown as dotted lines, which gives a quite good
approximation to the exact numerical result (solid lines). 

The condition (\ref{eq:Phi2}) leads to upper and lower limits to 
$u_-$, given by
\begin{eqnarray}
u_{\rm max}&\approx&{1\over2\beta_V}\left[
\left(\beta_V+1\right)\Gamma-\xi_1\right],\nonumber\\
u_{\rm min}&\approx& -{1\over2\beta_V}\left[
\left(\beta_V-1\right)\Gamma+\xi_1\right],
\label{eq:umax}
\end{eqnarray}
with
\begin{equation}
\Gamma=\gamma_{-0}+{\beta_V\tilde{E}^2_0\over2n_-(\beta_V-\beta_{-0})}
-{\xi_0-\xi_1\over1+h(\xi_1-\xi_0)\beta_V}.
\label{eq:Gamma}
\end{equation}
Using (\ref{eq:betapm}) the positron momentum $u_+$ can be expressed in terms of $u_-$, 
which leads to the same upper and lower limits for positrons as (\ref{eq:umax}).
For luminal waves with $\beta_V=1$, one has $u_{\rm max}\approx (1+\tilde{E}^2_0/n_-)/2$
and $u_{\rm min}\approx -1$. In this case oscillations skew strongly in the direction of 
wave propagation. For $\beta_V\to\infty$, (\ref{eq:umax}) reduces to $u_{\rm max}=
(\Gamma+u_{+0}+u_{-0})/2$ and $u_{\rm min}=-(\Gamma-u_{+0}-u_{-0})/2$. When $u_{\pm0}=0$, oscillations 
become symmetric with $u_{\rm max}=-u_{\rm min}=\Gamma/2$.
Since $u_{\rm max}\neq u_{\rm min}$, oscillating electrons and positrons have a net 
drift velocity $\sim (u_{\rm max}+u_{\rm min})/2\approx (\Gamma-\xi_1)/\beta_V$.
An accurate evaluation of the drift velocity is given in Sec. 3.3.

\begin{figure*}
\includegraphics[width=16cm]{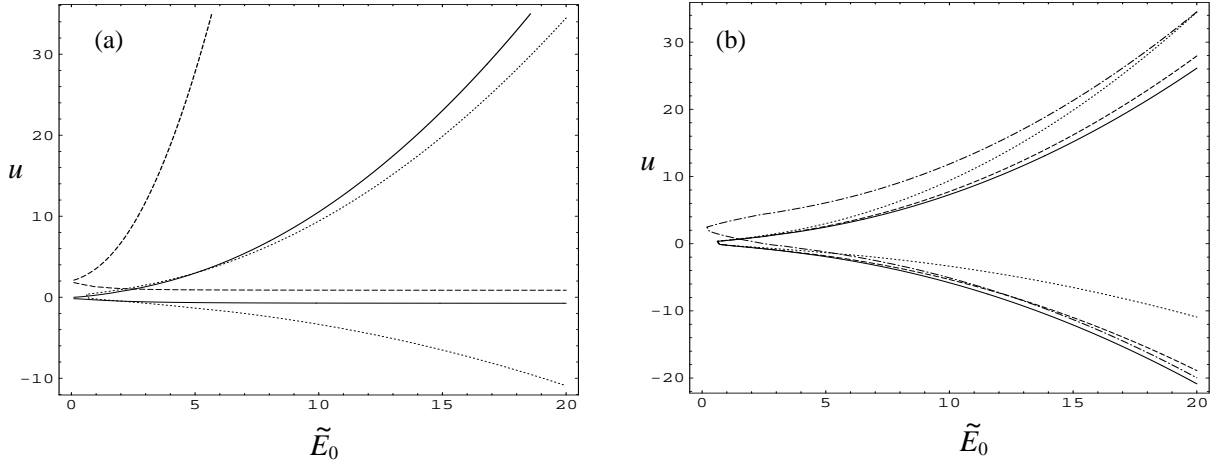}
\caption{Conditions for $E_\parallel=0$. (a) A luminal wave $\beta_V=1$, with
$n_-=5$, $j_{0\parallel}={\eta}_{GJ}=-1$. The solid and dashed lines correspond to 
$\beta_{\pm0}=0$ and $0.9$, respectively. Each pair of lines defines an upper and lower limits 
to $u$, between which $\Phi>0$. For comparison a superluminal wave with
$\beta_V=2$ and $\beta_{0\pm}=0$ is shown as a dot line. (b) Superluminal waves with 
different phase velocities $\beta_V=2$ (dotted), $10^3$ (solid), $10$ (dashed). 
We assume $\beta_{\pm0}=0$. The dash-dotted lines represent $\beta_V=10$ with 
$\beta_{0\pm}=0.9$.}
\label{fig:bounce}
\end{figure*}

Numerical solutions to $E_\parallel=0$ (i.e., $\Phi(\xi)=0$) are shown as contours in 
figure~\ref{fig:bounce}. We express $n_+$ in terms of $\beta_V$, $j_{0\parallel}$,
and $\eta_{GJ}$ using (\ref{eq:j-eta1}) and (\ref{eq:j-eta2}). 
In the subfigure on the left one assumes luminal waves 
with $\beta_V=1$. Each pair of lines defines an upper limit, $u_{\rm max}$, 
and a lower limit, $u_{\rm min}$, to a particle's momentum such that one has $\Phi>0$ 
for $u_{\rm min}<u_-<u_{\rm max}$. The similar upper and lower limits can be
obtained for positrons. Particles oscillate with their momenta confined
between these two limits. As $u_{\rm max}\gg |u_{\rm min}|$, the oscillations skew
strongly in the wave propagation direction. The dotted lines correspond 
to a superluminal wave $\beta_V=2$. As $\beta_V$ increases the system 
evolves toward symmetric oscillations as shown in the subfigure on the right.
A nonzero initial velocity ($\beta_{0\pm}>0$) also shifts the oscillations 
forward (dash-dotted lines). It should be emphasized that we intentionally choose a moderate $n_-$ 
here to illustrate that an oscillatory solution can exist even for a moderate 
pair density.

\subsection{Analytical formalism}

To seek an oscillatory solution one may define the oscillation periodicity (in units of
$1/\omega_{_{GJ}}$) 
\begin{equation}
\hat{T}={2\over\beta_V}\int^{\chi_b}_{\chi_a}{d\chi\over\Phi^{1/2}}=
-{2\over\beta_V}\int^{\xi_b}_{\xi_a}{d\xi\over\Phi^{1/2}(\xi)},
\label{eq:periodicity}
\end{equation}
where the subscripts $a$, $b$ label respectively the phases at which $\beta_-=\beta_{\rm min}$
and $\beta_-=\beta_{\rm max}$, respectively. The oscillation frequency is 
then given by $\omega=2\pi\omega_{_{GJ}}/\hat{T}$,
where $\omega_{_{GJ}}\approx (1.5\times10^{11}\,{\rm s}^{-1})P^{-1}_{0.1}B_8$.
Using the approximations (\ref{eq:g2}) an analytical solution for 
a LAEW can be derived. For $\beta_V\gg1$, using $\gamma_-(\xi)+g(\xi)\sim 2|\xi|/\beta_V$, 
one obtains the following approximation to (\ref{eq:Phi2}): 
\begin{equation}
\Phi(\xi)\approx2n_-\left(1-{\beta_{-0}\over\beta_V}\right)\left(\Gamma\pm{\xi_1-2\xi\over\beta_V\mp1}\right),
\label{eq:Phi3}
\end{equation}
where the upper and lower signs correspond to $\xi\geq\xi_1$ and $\xi<\xi_1$, respectively.
For $\beta_V\gg1$ one may expand (\ref{eq:Phi3}) in $1/\beta_V$ and substitute it for (\ref{eq:periodicity}), obtaining
\begin{equation}
\hat{T}\approx 4\left[{\beta_V\Gamma\over2n_-(\beta_V-\beta_{-0})}\right]^{1/2}\approx
{2\tilde{E}_0\over n_-}.
\end{equation}
A second approximation applies for $\tilde{E}^2_0/2n_-\gg\gamma_{0\pm}$, in which case
one estimates the frequency as
\begin{equation}
\omega\approx {\sqrt{2}\pi\omega_p\over4u^{1/2}_{\rm max}},
\label{eq:freq}
\end{equation}
where $\omega_p=(2n_-)^{1/2}\omega_{_{GJ}}$ is the plasma frequency of the pair plasma.
As an example, one has $\omega\approx 10^9\,{\rm s}^{-1}$ for $n_-=10^2$, $u_{\rm max}=10^6$ and 
$\omega_{GJ}=10^{11}\,{\rm s}^{-1}$. The oscillation frequency decreases as $u_{\rm max}$ 
increases, which can be understood as an increase in the effective mass of electrons or positrons.

\begin{figure}
\includegraphics[width=7.5cm]{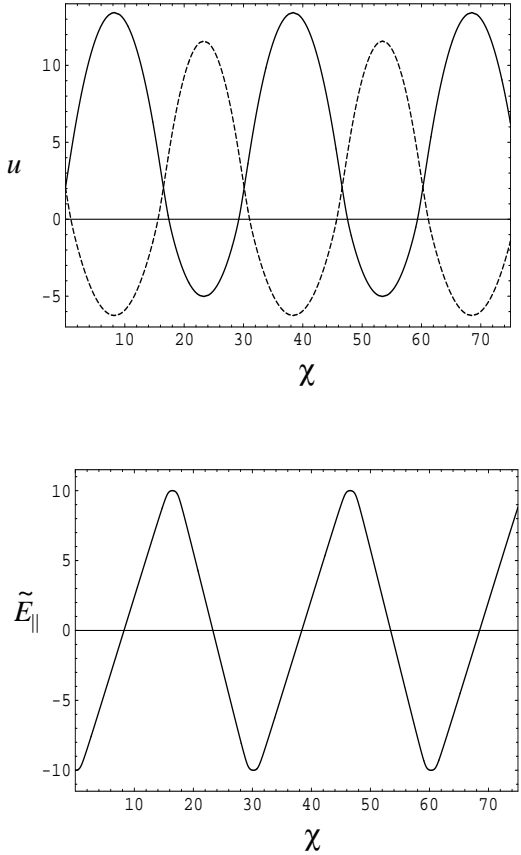}
\caption{Dimensionless momentum $u$ (upper) and electric field
$\tilde{E}_\parallel$ (lower) as functions of phase $\chi$. 
The dashed line corresponds to oscillations of positrons.
As there is no dissipation included, the amplitudes of the oscillations are determined
by $|\tilde{E}_0|$. Electrons oscillate between $u_{\rm min}\approx
-5.0$ and $u_{\rm max}\approx 13.4$, with a net drift velocity $\bar{u}\approx
3.6$. Oscillations of positrons (dashed) skew less in the positive direction than 
electrons. We assume $\tilde{E}_0=-10$, $\beta_V=5$, $\beta_{\pm0}=0.9$,
$j_{0\parallel}={\eta}_{GJ}=-1$, $n_-=5$.}
\label{fig:euchi}
\end{figure}

\begin{figure}
\includegraphics[width=8cm]{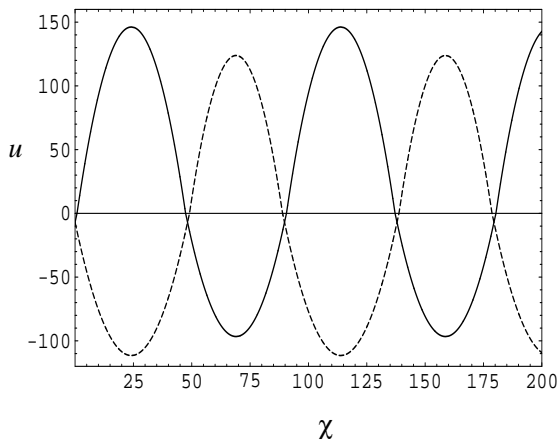}
\caption{As in figure~\ref{fig:euchi} but with $\beta_{\pm0}=-0.99$,
$\tilde{E}_0=-50$. The oscillation pattern is similar to 
figure~\ref{fig:euchi} but with a small backward shift.}
\label{fig:uchi}
\end{figure}

The condition for $\Phi>0$ is $|\xi|<\xi_b\approx-\xi_a\approx\beta_V\tilde{E}^2_0/4n_-$; this gives
$u_{\rm max}\approx \tilde{E}^2_0/4n_-\approx-u_{\rm min}$. Note that such symmetry in 
oscillation is a direct consequence of our assumption of large $\beta_V$ and
$\beta_{\pm0}=0$. The solution for $0\leq \chi<\chi_{_T}=\beta_V\hat{T}$  is found to be
\begin{eqnarray}
{\chi}\approx
\left\{
\begin{array}{l}
\displaystyle{
{\beta_V\over2n_-}\!\left(\Phi^{1/2}_0-\Phi^{1/2}\right)},
 \quad 0\leq\chi<{\textstyle{3\over4}}{\chi}_{_T},\\
\displaystyle{
{\textstyle{1\over2}}{\chi}_{_T}+{\beta_V\over2n_-}
\!\left(\Phi^{1/2}_0-\Phi^{1/2}\right)}, 
\, {\textstyle{3\over4}}{\chi}_{_T}\leq{\chi}<{\chi}_{_T},
\end{array}
\right.
\label{eq:chi}
\end{eqnarray}
where $\Phi$ is given by (\ref{eq:Phi3}), $\Phi_0=\Phi(1)$,
$\Phi_{a,b}=\Phi(\xi_{a,b})$. The electric field is obtained as
\begin{equation}
\tilde{E}_\parallel\approx\left\{
\begin{array}{ll}
\displaystyle{\tilde{E}_0+{2n_-\over\beta_V}{\chi}
},&0\leq{\chi}<{\chi}_{_T}/2,\\
\displaystyle{\tilde{E}_0+{2n_-\over\beta_V}({\chi}_{_T}
-{\chi})
}, 
& {\chi}_{_T}/2\leq{\chi}<{\chi}_{_T}.
\end{array}
\right.
\label{eq:E}
\end{equation}
The electric field displays a sawtooth wave form, which can be understood 
qualitatively in terms of the extreme relativistic limit. In this limit positrons and electrons 
are accelerated in opposite directions, giving rise to 
a current $|j_\parallel|\sim 2|\beta_-|n_-\gg |j_{0\parallel}|\sim 1$. 
Thus, the electric field is $E_\parallel\propto \pm |j_\parallel|/\beta_V\sim\pm 2n_-\chi/\beta_V$ 
with $|\beta_-|\sim1$, which reproduces the sawtooth wave form given by (\ref{eq:E}). 
An example of a numerical 
integration of (\ref{eq:EqMotion4}) and (\ref{eq:waveE4}) is shown in figure \ref{fig:euchi} for
pairs with an initial, forward velocity. Figure~\ref{fig:uchi} shows oscillations for
particles with an initial velocity toward the star.  
Although our analytical solution is obtained for $n_-\gg1$, here we again 
choose a moderate $n_-$ in the numerical calculation to show that our oscillatory 
solution is also valid for $n_\pm\sim1$.
The wave form is similar to that predicted from the numerical model
\citep{letal05}. The characteristics of the oscillations, including
the periodicity and amplitude, are independent of the sign of $\tilde{E}_0$ and 
are not sensitive to the initial conditions $\beta_{0\pm}$ provided that $|u_{0\pm}|\ll u_{\rm
max}$. As no radiative loss, pair production nor 
wave damping is included, the wave amplitude remains constant. In the figure 
we assume an initial electric field much lower than the typical vacuum 
field $\tilde{E}_{\rm max}\sim 3\times10^6(B/B_c)^{1/2}
P^{-1/2}_{0.1}$, where $B_c\approx 4.4\times10^9\, \rm T$. 
In practice, the initial field should be near the pair creation 
threshold. Assuming the pair production threshold to be $\gamma_{th}$, one
has $\tilde{E}_0\approx3\times10^4(n_-/5)^{1/2}(\gamma_{th}/10^6)^{1/2}$
(see discusssion in Sec. 5). Since $|u_{\rm min}|<|u_{\rm max}|$, 
the oscillating particles have a net forward flow velocity.

\subsection{Drift motion}

The drift momentum can be obtained by averaging $u_-$ over one period ($\hat{T}$):
\begin{eqnarray}
\bar{u}&=&{2\over\beta_V\hat{T}}\int^{\chi_b}_{\chi_a}u_-d\chi
\approx
{2\Gamma\over3\beta_V}\left(1-{3\xi_1\over4\Gamma}\right),
\label{eq:ubar}
\end{eqnarray}
where we expand $\Phi$ on $1/\beta_V\ll1$ and change the integration variable to
$d\chi=-d\xi/\Phi^{1/2}(\xi)$. Terms of order $|j_{0\parallel}|/n_\pm\ll1$ and $|\eta_{_{GJ}}|/n_\pm\ll1$
are ignored. Then both electrons and positrons are dragged along in the wave at the same
drift velocity $\beta_D=\bar{u}/(1+\bar{u}^2)^{1/2}$. For $\tilde{E}^2_0/2n_-\gg\gamma_{\pm0}$ one has
\begin{equation}
\bar{u}\approx {\tilde{E}^2_0\over3n_-\beta_V}\approx{4u_{\rm max}\over3\beta_V}.
\label{eq:ubar2}
\end{equation}
The drift velocity decreases as the wave phase speed increases.
In the limit $\beta_V\to\infty$ oscillations are purely temporal. The upper and lower
limits to the particle's momentum are $u_{\rm max}=(\Gamma+u_{+0}+u_{-0})/2$ 
and $u_{\rm min}=-(\Gamma-u_{+0}-u_{-0})/2$. The drift velocity (\ref{eq:ubar}) reduces
to $\bar{u}=(u_{+0}+u_{-0})/2$. For $u_{\pm0}=0$, particles oscillate symmetrically between 
$u_{\rm min}\approx-u_{\rm max}\approx \tilde{E}^2_0/4n_-$ and $u_{\rm max}$.
It is interesting to note that the proportionality $1/\beta_V$ in (\ref{eq:ubar2}) is
similar to that predicted from the single-particle treatment \citep{r92b}. 
However, since the single-particle formalism does not include the feedback effect of 
particles on the wave, it predicts a low drift velocity $\beta_D\approx1/\beta_V$.
Our extact treament shows that the drift motion can be highly relativistic with 
$\gamma_D\equiv1/(1-\beta^2_D)^{1/2}\approx |\bar{u}|\gg1$. 
As a result, the LAEW can drive a relativistic outflow of particles 
even when the particles are initially at rest. Thus, the oscillating gap can supply
relativistic pairs to the pulsar wind. 

An implication of such drift motion is that polar cap overheating can be avoided
and so, the model satisfies the observational constraint on
thermal X-rays from the polar caps. The observed relatively low fluxes of 
thermal X-rays imply either that the flux of particles that impact
on the polar cap is much lower than the GJ flux $\sim|\rho_{GJ}|c$ or that 
acceleration of returning particles is insignificant. The latter can be ruled out
as the returning particles must be subject to the same strong accelerating 
electric field that accelerates forward moving particles.
In the oscillatory model, since particles are dragged forward by the LAEW and escape
to infinity, few particles are reflected back to the star; oscillating particles do not 
impact on the polar cap if the oscillating region is located
a distance $>c\hat{T}/\omega_{GJ}\approx\lambda/\beta_V\sim 0.9\,\rm m$
for $B=10^8\,\rm T$ and $P=0.1\,\rm s$. So, the model can satisfy the observational 
limit to the thermal X-ray observations from the polar cap. 
It is worth commenting that by constrast, \citet{rs75}'s `vacuum sparks' model, 
which is instrinsically time dependent, predicts a much larger thermal X-ray flux 
than the observational limit.

\subsection{Currents}

\begin{figure}
\includegraphics[width=7.5cm]{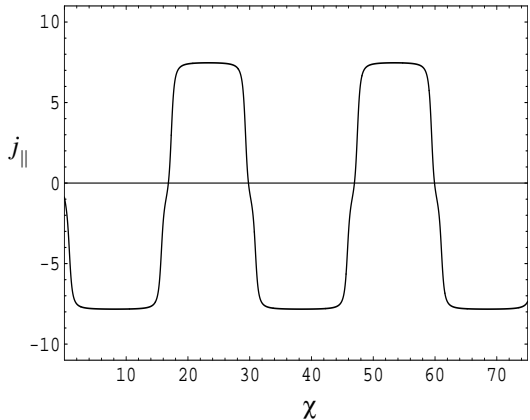}
\caption{Current $j_\parallel$  vs $\chi$.
The parameters are as in figure \ref{fig:euchi}. The square wave form can be understood
as that in each oscillation particles stay most time in the relativistic regime in which
$\beta_\pm\sim 1$.}
\label{fig:j}
\end{figure}

The current $j(\chi)$ can be derived using (\ref{eq:chi}). Assuming
$\beta_V\gg1$, we have
\begin{eqnarray}
j_\parallel&\approx& \left(1-{\beta_{+0}\over\beta_V}\right)\beta_+n_+-
\left(1-{\beta_{+0}\over\beta_V}\right)\beta_-n_-\nonumber\\
&&+{1\over\beta_V}\left(\beta^2_+n_+-\beta^2_-n_-\right).
\label{eq:j}
\end{eqnarray}
For half the phase of an oscillation, the
electrons and positrons are accelerated in opposite directions,
and then these directions reverse, with only a brief phase in which the
motion is nonrelativistic. Hence, one has 
$\beta_+\sim -\beta_-\sim1$ for nearly all phases. The final term
in Eq (\ref{eq:j}) is generally small, and on neglecting it,
the current is  $j_\parallel\approx \pm 2n_-$ except for the short phase 
where it switches sign. Thus, the oscillating current has a square wave form with
an amplitude $|j_{\rm max}|\sim 2n_-\gg |j_{0\parallel}|$.
A numerical calculation of $j_\parallel(\chi)$ is shown in figure \ref{fig:j}.
Note that as we consider only electrostatic waves, so that the oscillating current 
induces electric fields only; there is no oscillating magnetic field. 

The average current is given by
\begin{eqnarray}
\bar{j}_\parallel\approx \bar{\beta}_+n_+-\bar{\beta}_-n_-\approx \bar{\beta}(n_+-n_-),
\label{eq:mj}
\end{eqnarray}
where $\bar{\beta}_\pm$ is the mean velocity, i.e., drift velocity.  The second approximation is derived
for $n_\pm\gg1$ and hence $\bar{\beta}_+\approx\bar{\beta}_-$. The mean charge density
is given by $\bar{\eta}=\eta_{GJ}+(\bar{j}_\parallel-j_{0\parallel})/\beta_V$. The system tends to settle
into a state where $\bar{\eta}=\eta_{GJ}$ and $\bar{j}_\parallel=j_{0\parallel}$. 

\subsection{Wave dissipation}

So far we neglect dissipation in obtaining our solution for LAEW.
This is justified if damping (or growth) is weak in the sense that the change in the LAEW 
in an oscillation period can be treated as a perturbation. Damping occurs through radiative 
losses, which include curvature radiation, RICS, two-stream instability, and 
linear acceleration emission (LAE) \citep{m78}. In the first and second mechanisms, energy
losses occur through pair creation. In the third mechanism, plasma instability
can arise from counterstreaming of electrons and positrons in oscillations.  
In the forth mechanism, particles accelerated in LAEW directly emit electromagnetic radiation, which 
has close analogy to synchrotron radiation or inverse Compton scattering. It can be
shown that damping due to these radiation processes is indeed weak, with the damping time 
being much longer than the wave period and generally longer than
the light-crossing time over the gap. Although in principle one 
may determine the damping from (\ref{eq:EqMotion})--(\ref{eq:curlB}), averaged over 
a wave period \citep{aetal77}, here in estimating the effect of wave damping, we adopt 
a different approach in which the wave damping is related to energy 
losses by a single particle. 

To estimate the damping time we consider how the total wave energy density evolves.
The total wave energy density can be written as a sum of the electric energy density,
$U_E=\varepsilon_0 E^2/2$, and the energy density associated with particle oscillations,
$U_p=m_ec^2\gamma_{max}(n_++n_-)N_{GJ}$. Using $\gamma_{max}\approx u_{max}\approx
e^2E^2/(m^2_ec^2\omega_{GJ})$ (cf. Sec 3.2), one finds $U_p\approx U_E$.
Let the average energy loss for a single particle be $\langle\dot{\gamma}\rangle$,
where the average is taken over one wave period. The typical damping time scale
can be estimated from 
\begin{equation}
\tau\approx -{U_E+U_p\over (n_++n_-) N_{GJ}\langle\dot{\gamma}\rangle m_ec^2}\approx
-{2\gamma_{max}\over \langle\dot{\gamma}\rangle}.
\end{equation}
As an example, for curvature radiation, one has $\dot{\gamma}_{\rm curv}=-(2r_ec/3R^2_c)\gamma^4$, where
$r_e\approx 2.8\times 10^{-15}\,\rm m$ is the classical electron radius. 
Using the average $\langle\gamma^4\rangle\approx0.4\gamma^4_{max}$ in the 
limit $\beta_V\gg1$, one obtains $\langle\dot{\gamma}_{\rm curv}\rangle\approx 
-0.4(2r_ec/3R^2_c)\gamma^4_{max}$. For $R_c\approx 3\times10^5\,\rm m$ and 
$\gamma_{max}=10^6$, one has $\tau\approx7.5R^2_c/(r_ec\gamma^3_{max}) \approx 0.8\,\rm s$. 
Thus, the damping time is much longer than the wave period ($2\pi/\omega$) and 
considerably longer than the light-crossing time over the gap, denoted by $\tau_g$.
In general, one has $\tau_g\leq R/c\approx3\times10^{-4}\,\rm s$. A similar estimate 
can be obtained for other three processes and it can be shown that
our assumption $\tau\gg 2\pi/\omega$ is valid and that in general $\tau>\tau_g$. 
For fast pulsars, curvature radiation may become efficient, as a result of 
a smaller curvature radius, and the damping time 
may become comparable or even shorter than $\tau_g$, but it is still much longer
than the wave period. Similarly, for high magnetic field pulsars ($B>B_c$) 
with hot polar caps, resonant inverse Compton scattering \citep{s95} (cf. Sec 4.2) can 
be efficient and can also lead to $\tau\leq \tau_g$.  

\section{Pair creation}
\label{pair-production}

In this section we discuss pair creation occurring in a LAEW. The dominant 
pair production process is single photon decay in superstrong magnetic fields. 
We consider two main emission processes that produce pair-producing
photons: curvature radiation and resonant inverse Compton scattering (RICS)
\citep{s95,l96}. The latter is inverse Compton scattering in cyclotron resonance, involving
scattering of thermal photons from the star's surface by relativistic 
electrons or positrons. Although other processess such as nonresonant inverse Compton
scattering may also contribute to pair production, we only focus these processes. 
How pair creation affects the wave depends on
the ratio of the pair production free path, $\lambda_p$, and 
the wavelength, $\lambda$. For $\lambda_p/\lambda\gg1$, a photon 
travels many wavelengths before it decays into a
pair. Pair injection can be treated as nonoscillatory, with the injection rate
derived from a sum over the pair production during many oscillations. In the opposite limit
$\lambda_p/\lambda\ll1$, pair creation is locked in oscillations: pairs are injected
at particular phase during each oscillation. Here we only discuss the first limit  
as it is applicable for typical pulsars.

\subsection{Curvature radiation}

The free path for pair production can be written as 
$\lambda_p=\Delta s_i+\Delta s_p$, where $\Delta s_i$ is the
characteristic length for emission of a photon at energy
$>2m_ec^2$ and $\Delta s_p$ is the path length that the photon
needs to travel before decay into a pair. For curvature radiation one can
show that the former is much shorter than the latter.
To estimate $\Delta s_i$ one writes the production rate, $dn^{(\pm)}_{\rm ph}/dt$,
of forward ($+$) and backward ($-$) propagating curvature photons at the energy
$\varepsilon_c$ as a ratio of the radiation power, $P_{\rm curv}\sim -\dot{\gamma}_{curv}$,
to the characteristic energy, $\varepsilon_c$, of curvature photons emitted by a relativistic
electron moving along a curved field line with a curvature radius, $R_c$.
Using $dn^{(\pm)}_{\rm ph}/dt\approx -\dot{\gamma}_{\rm curv}/\varepsilon_c$,
where $\varepsilon_c\approx(3\lambda_c/2R_c)\gamma^3=2P^{-1/2}_{0.1}(\gamma/10^6)^3$,
and $\lambda_c=\hbar/m_ec\approx3.86\times10^{-13}\,\rm m$ is the Compton wavelength,
one obtains
\begin{equation} 
{dn^{(\pm)}_{\rm ph}\over dt}\approx {\alpha_fc\over R_c}\gamma,
\end{equation}
where $\alpha_f\approx1/137$ is the fine constant.
Since $dn^{(+)}_{\rm ph}/dt\approx dn^{(-)}_{\rm ph}/dt$ for $z\ll R$, the forward 
and backward components are approximately symmetric.
For a dipole magnetic field, the curvature radius is given by
$R_c=(4/3)(crP/2\pi)^{1/2}\approx2.9\times10^5(r/R_0)^{1/2}P^{1/2}_{0.1}
\,{\rm m}$. We have
\begin{equation}
\Delta s_i\approx {R_c\over \alpha_f\gamma_{th}}\approx 13.7P^{1/2}_{0.1}
\left({10^6\over\gamma_{th}}\right)\,{\rm m}.
\end{equation}
A photon with energy $\varepsilon_\gamma\sim\varepsilon_c>2$ needs to travel a further distance 
before being converted to a $e^\pm$ pair. This distance can be estimated as
follows. The opacity of a photon in a strong magnetic field is
a function of $\psi=0.5\varepsilon_B\varepsilon_\gamma\sin\theta_{\gamma B}$, where
$\theta_{\gamma B}$ is the propagation angle of the photon and $\varepsilon_B=B/B_c$ \citep{e66}. Generally, pair creation
requires $\psi\sim 1/15$. A pair is produced when the opacity reaches unity. Using $\theta_{\gamma B}\sim 
\Delta s_p/R_c$, one has $\Delta s_p=2\psi R_c/\varepsilon_B\varepsilon_c$ to produce 
one pair. Since the maximum $\gamma$ is limited by radiation-reaction, denoted 
by $\gamma_R$, one may obtain a lower limit to the free path,
\begin{equation}
\lambda_p={R_c\over\varepsilon_{c,R}}
\left[
{\varepsilon_{c,R}\over\alpha_f}\left({3\lambda_c\over2R_c}\right)^{1/3}
+{2\psi\over\varepsilon_B}\right],
\end{equation}
where $\varepsilon_{c,R}=(3\lambda_c/2R_c)\gamma^3_R$.
The right-hand side is $\sim 500\,\rm m$ for $R_c=3\times10^5\,\rm m$, 
$\psi=0.01$, $\varepsilon_B=0.1$, $\varepsilon_\gamma=\varepsilon_{c,R}=10^2$. 
Using the parameters in (\ref{eq:lambda}) one has $\lambda<\lambda_p$ for $\beta_V<5\times10^2$.

\subsection{Resonant inverse Compton scattering}

Similarly, one may estimate $\lambda_p$ for RICS. For thermal photons 
at energy $\Theta<1/\gamma$, one can ignore the Klein-Nishina effect;
the production rate of the scattered photon at energy $\varepsilon_s\sim\varepsilon_B\gamma$ is
\citep{s95,l96}
\begin{equation}
{dn^{(\pm)}_{\rm ph}\over dt}\approx 
{9x^{(\pm)}\Theta\varepsilon_B\over8\pi^2\gamma^2}{c\sigma_{\rm eff}\over\lambda^3_c},
\end{equation}
where $\Theta=1.7\times10^{-4}(T_s/10^6\,{\rm K})$ is the normalized temperature 
$T_s$ of the polar cap, $\sigma_{\rm eff}\approx 3\pi\sigma_T/4\alpha_f$ 
is the effective cross section of RICS, and 
$x^{(\pm)}=-\ln[1-\exp(-\varepsilon_B/\Theta\gamma(1\mp\beta\cos\theta_m))]$ with 
$\theta_m$ the maximum propagation angle of the incoming photon \citep{d90}. 
Since $x^{(-)}>x^{(+)}$, pair production by particles moving toward
the star is more efficient than particles moving away from the star \citep{hm98}.
For both cases, the rate increases with decreasing $\gamma$. However, the process becomes less efficient
at low energy as the cyclotron resonance condition becomes difficult to satisfy 
($x^{(\pm)}$ decreases exponentially when $\gamma$ is too low for the condition to be
satisfied). This leads to an estimate of $\Delta s_i$:
\begin{equation}
\Delta s_i\approx 10^4\epsilon^{1/3}\left({T_s\over10^6\,{\rm K}}\right)^{-2/3}
\left({x_\pm\over0.5}\right)^{2/3}B^{-4/3}_8P^{1/6}_{0.1}\,{\rm m}.
\end{equation}
We write $\gamma$ as a fraction $\epsilon\ll1$ of the maximum
potential drop across the polar cap. Since $\Theta<1/\gamma$, one has
\begin{equation}
\Delta s_p={2\psi R_c\over\gamma\varepsilon_B}>2\psi R_c{\Theta\over
\varepsilon_B}.
\end{equation}
For $\Theta=1.7\times10^{-4}$ and $\varepsilon_B=0.1$, one has $\Delta s_p\approx 23\,{\rm m}$.
One concludes that for a moderate $\beta_V>1$, $\lambda_p\gg\lambda$ applies to RICS.
For $\gamma>1/\Theta$, the scattering is in the Klein-Nishina regime, which is not discussed
here.  

\subsection{Pair injection}

Since for $\lambda_p\gg\lambda$, one may regard $dn^{(\pm)}_{\rm ph}/dt$ as a constant,
where $\gamma$ is replaced by its average (over the period).
One may write the source term in (\ref{eq:ContinuityEq2}) as
\begin{equation}
Q={\textstyle{1\over2}}(\alpha_+N_++\alpha_-N_-),
\label{eq:source}
\end{equation}
where $\alpha_\pm=\langle dn^{(+)}_{\rm ph}/dt\rangle+
\langle dn^{(-)}_{\rm ph}/dt\rangle$ is a 
constant. It should be noted that inclusion of higher generations of
pairs is necessary in conventional treatments of pair creation in a pair formation front, 
especially for curvature radiation.  Inclusion of higher generations would effectively 
increase $\alpha_\pm$ in (45). However, the 
assumption that the free path for pair production is much longer than the 
wavelength remains valid.  

Assuming $\beta_V\gg1$, integration of (\ref{eq:ContinuityEq2}) 
results in an exponential growth
\begin{equation}
N_\pm\approx n_\pm\exp\left(
{\bar{\alpha}_++\bar{\alpha}_-\over2\beta_V}\chi\right),
\label{eq:injection}
\end{equation}
where $\bar{\alpha}_\pm=\alpha_\pm/\omega_{GJ}$.
One can show that for $\chi\sim \beta_V\omega_{GJ}t$ as $\beta_V\gg1$,  
Eq (\ref{eq:injection}) reduces to $N_\pm\propto \exp[-(\alpha_++\alpha_-)t/2]$,
which reproduces the numerical result in \citet{letal05}. 

We envisage pair creation as being only a minor perturbation except when 
the large-amplitude oscillation is being set up, as described in the purely
temporal case by Levinson et al. (2005). Here one may derive the condition 
under which the effect of the source function in the equation of
motion (\ref{eq:EqMotion2}) can be ignored. This effect can be
characterized by a parameter $\delta_\pm\equiv \tilde{Q}/2\beta_V\tilde{N}_\pm$, where
we assume $\beta_V\gg1$. The pair production has only a minor effect if
$\delta_\pm\ll 1$. Using (\ref{eq:source}) and (\ref{eq:injection}), this condition 
can be writtten in the form $\delta_\pm=(\alpha_+n_++\alpha_- n_-)/(4\beta_V\omega_{GJ}n_\pm)\ll1$.
As an example, for curvature radiation 
one has $\delta_\pm\sim 2^{m_g-1}\alpha_f c\gamma/2R_c\beta_V\omega_{GJ}\ll1$, 
where one assumes there are $m_g$ generations of pairs. 

\section{Low-density limit}
\label{low-density}

The low-density regime is applicable if there are initially insufficient charges to 
provide the GJ charge density. One can show that in this limit both monotonic and oscillatory 
solutions exist, with the former corresponding to rapid acceleration of charged particles 
from the surface. 

\subsection{Monotonic acceleration}

\begin{figure}
\includegraphics[width=7.5cm]{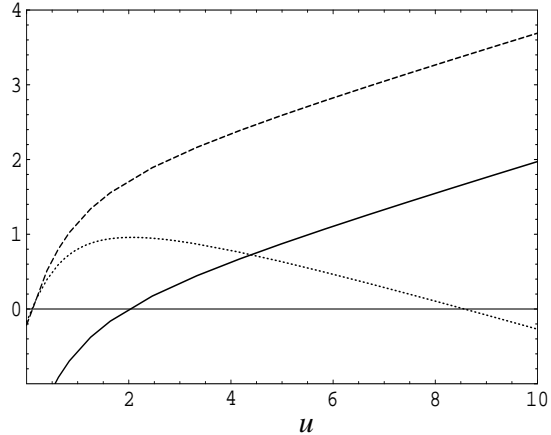}
\caption{Plot of $\Phi$ as a function of $u$ in the low-density limit
for $(\eta_{GJ},j_{0\parallel})=(-1,-1.1)$ with $u_{-0}=2$ (solid),
$(\eta_{GJ},j_{0\parallel})=(-1,-1.1)$ with  $u_{-0}=0.1$ (dashed) and 
$(\eta_{GJ},j_{0\parallel})=(-1,-0.9)$ with $u_{-0}=0.1$ (dot).
The third case only permits an oscillatory solution with $u$ being limited by 
$\leq u_{\rm max}\approx 8.5$. The initial electric field is assumed to be $\tilde{E}_0=0$.
}
\label{fig:mono}
\end{figure}

A relevant example of a low density corresponds to a space-charge-limited flow
(SCLF) from the surface, which has so far been discussed only in the context of the 
steady state \citep{as79,ms94,hm98}.  The case of a vacuum-like field
is applicable for pulsars with $\bOmega\cdot\bB<0$ when ions are tightly bound to 
the surface \citep{ml07}. This case was considered in \citet{letal05} and is not discussed
here. We consider an outflow of electrons from 
the polar cap with $\bOmega\cdot\bB>0$. The initial electric field at the surface
is assumed to be small. Assuming $n_+=0$, one obtains
\begin{equation}
\Phi=\tilde{E}^2_0+2\Bigl[(\gamma-\gamma_0)\eta_{GJ}-(u-u_0)j_{0\parallel}\Bigr],
\label{eq:Phi4}
\end{equation}
where (\ref{eq:j-eta2}) is used. A solution for $\Phi>0$ is
$j_{0\parallel}<\eta_{GJ}$ for $u>u_0\geq0$. As shown in figure \ref{fig:mono},
$u$ only has a lower bound, corresponding to the positive part of 
the solid and dashed lines, and in these cases electrons can be accelerated monotonically
outward. When $j_{0\parallel}>\eta_{GJ}$ (dotted line), $u$ has both upper and lower bounds
and the solution must be oscillatory with the upper bound being well below the 
pair production threshold. Such low amplitude oscillatory solution was also discussed
recently \citep{b07}. It should be pointed out here that the effect of 
the conducting wall is not considered but can become dominant in determining 
the acceleration (see further comments in Sec 5.2). For $\beta_V\gg1$ one has $d\xi\approx -\beta_V du$
in (\ref{eq:EqMotion4}), which gives rise to a monotonic solution for the electric field:
\begin{equation}
\tilde{E}_\parallel\approx \tilde{E}_0-{\delta\eta\over\beta_V}\chi,
\label{eq:E1}
\end{equation}
with $\delta\eta=\eta_{GJ}-j_{0\parallel}>0$ and $\tilde{E}_0<0$. 
The electron's momentum increases according to
\begin{equation}
u_-\approx u_{-0}+{|\tilde{E}_0|\over\beta_V}\chi+{\delta\eta\over2\beta^2_V}\chi^2.
\label{eq:u1}
\end{equation}
A special case is an initial phase $\chi=0$ chosen corresponding to $t=0$ and $z=0$, 
i.e., at the surface where one has $\tilde{E}_0\approx0$. We envisage that 
$\tilde{E}_\parallel$ grows to reach the value at which pair production occurs
and the oscillatory phase takes over with $\tilde{E}_0$ in Sec 3 replaced by
the threshold $\tilde{E}_\parallel$. The acceleration discussed 
here is different from the usual SCLF models; the electric field is 
predominantly inductive.

In the temporal case, an electric field increases linearly with time and thus, 
the 4-velocity of the particles varies quadratically with time: 
$u_-\propto \delta\eta\tilde{t}^2/2$ with $\tilde{t}=\omega_{GJ}t$.
The time for an electron to be accelerated to the pair production
threshold $\gamma_{th}$ is
\begin{equation}
t={2\over\omega_{GJ}}\left({\gamma_{\rm th}\over|\delta\eta|}\right)^{1/2}.
\label{eq:t}
\end{equation}
The time (\ref{eq:t}) should be limited by the travel time of a particle across
the acceleration region. Assuming the region to have a size of $\Delta L$, 
the limit gives rise to a condition $t<\Delta L/c$ for a pair cascade to occur.

In general, the relative importance of acceleration due to inductive 
and non-inductive (potential) electric fields can be determined from 
$(u_-)_{\rm inductive}/(u_-)_{\rm static} \sim \beta^2_V$. For $\beta^2_V\gg1$,
inductive acceleration is dominant. The acceleration 
can be regarded as purely electrostatic only in the special
case $\beta_V\to 0$, which is discussed in 5.2. 

\subsection{Comparison with the steady-state models}

The steady state can be regarded as a limit $\beta_V\to0$ in which
the parallel electric field is purely static. For $\beta_V\to0$ the phase 
depends on spatial coordinates only and can be written as
$\chi=-\omega_{GJ}z/c\equiv-\tilde{z}$. 
Substituting (\ref{eq:Phi4}) into (\ref{eq:EqMotion4}) and using 
$d\xi\approx d\gamma$, one obtains 
\begin{equation}
\tilde{E}_\parallel=\tilde{E}_0-\delta\eta\tilde{z},
\label{eq:E2}
\end{equation}
where $\tilde{E}_0<0$.  The monotonic solution (\ref{eq:E2}) can be 
regarded as an oscillatory solution in the long period limit $\hat{T}\to\infty$.
From the current-charge invariant (\ref{eq:j-eta1}) one has 
$\eta(0)-\eta_{GJ}=-(\beta_{-0}n_-+j_{0\parallel})/\beta_V$.
Since $\eta(0)-\eta_{GJ}$ must remain finite, the limit $\beta_V\to0$ implies
that the initial current is $j_\parallel(0)\approx-\beta_{-0}n_-$, which matches the constant 
current $j_\parallel(0)=j_{0\parallel}$. The electron's momentum is derived as
\begin{equation}
u_-=u_{-0}+|\tilde{E}_0|\tilde{z}+{\textstyle{1\over2}}\delta\eta\tilde{z}^2.
\label{eq:u2}
\end{equation}
As for (\ref{eq:E1}), for electrons to be accelerated outward one must 
have $\delta\eta>0$. That the same condition ($\delta\eta>0$) is required in both cases
is hardly surprising. In (\ref{eq:E1}), one has $d\xi\sim -\beta_V du<0$
and $\chi>0$, while in (\ref{eq:E1}) one has $d\xi\sim du>0$ but 
$\chi<0$. If $\tilde{E}_0\sim0$, the electron's momentum increases with $\tilde{z}$
quadratically, $u_-\sim \delta\eta\tilde{z}^2/2$.

In the conventional SCLF models \citep{as79,hm98}, $E_\parallel<0$ is obtained 
with $\delta\eta<0$, by imposing an upper boundary, usually located at the PFF,  
where $E_\parallel=0$, and a conducting surface of the side wall of the open field line 
region; in these models $j_{0\parallel}$ is then determined locally 
by these boundary conditions. The basic assumption in the SCLF models is the nonconstancy 
of $\delta\eta$ along flow. This means that if one sets $\delta\eta=0$ initially, a 
nonzero $\delta\eta\neq0$ develops along the flow inducing a parallel electric field.
Two effects that lead to $\delta\eta<0$ have been considered in the literature,
including field line curvature, corresponding to the field lines curving
toward the rotation axis, and frame dragging \citep{mt92}. The latter dominates
near the star; the effective angular velocity, so is the GJ density, is reduced 
by a factor $(1-k_g(R/r)^3)<1$ as compared to that observed in a flat space at 
infinity, where $k_g=2GI/(c^2R^3)\approx0.15I_{38}$, $z<R=10^4\,\rm m$,
and $I_{38}=I/(10^{38}\,{\rm kg}\,{\rm m}^2)$ is the moment
of inertia of the star \citep{mt92}.
If one assumes $\delta\eta=0$ initially at the surface, one has
$\delta\eta=3k_g\eta_{GJ}z/R<0$ for $\eta_{GJ}<0$.

Eq (\ref{eq:E2}) and (\ref{eq:u2}) are similar to 
the result derived by \citet{s97} based on a generic SCLF model in which 
no specific local boundary condition is imposed. 
When $j_{0\parallel}$ is treated as a free parameter, for initially $\delta\eta=0$,
$\delta\eta>0$ is required to produce $E_\parallel<0$. This can occur
only on the curving-away (from the rotation axis) field
lines along which $|\bOmega\cdot\bB|$ decreases \citep{s97,m99}.
A major problem with this scenario in the context of the steady-state
limit is that the growth in $|E_\parallel|$ is unstoppable \citep{s97,m99}.
However, such run-away growth does not occur in our oscillatory model because
pair creation ultimately leads the system to switch to an oscillatory
phase, as discussed in Sec. 3.
For $\delta\eta<0$, Eq (\ref{eq:u2}) implies an oscillatory (in space)
solution similar to that found previously \citep{ms94,s97}.
When the acceleration region extends to $>R(R/R_{LC})^{1/2}$, the effect of
the conducting side wall, at which $E_\parallel=0$, becomes important. 
When such effect is included acceleration of outflowing electrons is
possible at $>R(R/R_{LC})^{1/2}$ even when $\delta\eta<0$ (provided that an 
electric field arising from such effect dominates over that from $\delta\eta<0$) \citep{s97}. 

\section{Conclusions and discussion}

We present an oscillatory polar gap model, in which the system initially undergoes a 
low-density phase, involving rapid acceleration of particles to ultra high 
energy, initiating a pair cascade. The system evolves to an oscillatory phase.
The oscillations are treated as a superluminal, large amplitude electrostatic wave that propagates
along the magnetic field. The charge continuity equation implies a
current-charge invariant ($j_\parallel-\beta_V\eta={\rm const}$)
that is independent of pair creation. As a result, the phase velocity $\beta_V$ is 
no longer a free parameter and can be written in terms of the initial velocity and
density of the plasma. It is shown that only the superluminal case $\beta_V>1$ is relevant here.
An analytical formalism for LEAWs is derived in the high-density
regime in which the pair density is higher than the GJ density. 
We ignore wave damping in our analytical solution. Neglecting 
damping is justified as the typical damping time due to 
energy losses through radiation is much longer than the wave period.
In most cases, the damping time is also longer than the light-crossing 
time over the gap.  

The model predicts an outflow of relativistic pairs due to particles
being dragged along in LAEW. Such feature is needed to avoid overheating of the 
polar cap. Outflowing pairs would contribute to the pulsar wind.
Pairs oscillate with a net drift velocity directed along the 
magnetic field, producing a current that oscillates about the global constant current $j_0$. 
The amplitude of the oscillating current is larger than the global current by a 
large factor that is of order of magnitude the ratio of the pair density to the GJ density. 
The wave form of an inductive electric field is
characterized by a triangular shape, which can be understood as the current being 
nearly constant except for a brief period during which it switches sign. 
The basic features of the oscillations are not sensitive to the initial 
conditions including the electron's or positron's initial velocity. 

There are two possiblities for particle acceleration in the initial 
phase that leads to oscillations: (1) a vacuum-like initial electric field, which 
may be applicable for the polar cap where charges are tightly bound to the 
surface, and (2) SCLF, in which there is an ample supply of charges.
The first case was discussed in \citet{letal05}. Here we consider specifically the 
SCLF case where an initial electric field appears as a result of an imbalance between the 
charge density and the GJ density with the latter mimicking the positive 
background charges. Electrons are accelerated monotonically in 
the electric field that increases linearly with the phase $\chi$. 
Since $\chi$ comprises both temporal and spatial variables, such particle acceleration 
arises from a mixture of inductive and non-inductive effects.
An interesting limit is $V\to \infty$, in which the electric field becomes
purely inductive. Qualitatively, the usual steady-state theory can be reproduced in the
limit of a zero phase speed. In this limit, the system is time independent
and the acceleration occurs at a specific spatial location. 
By contrast, acceleration due to an inductive field can occur 
everywhere in the region concerned.

An implication of the oscillatory model is the prediction of plasma instability 
arising from counterstreaming of electrons and positrons; in each oscillation electrons 
and positrons are accelerated in opposite direction and
such counterstreaming provides an ideal condition for two-stream instability which 
may be directly relevant for pulsar radio emission \citep{vm07}. 
Although various forms of streaming instability have been discussed in 
connection with the radio emission in conventional models, the growth rate is 
generally too low to be effective, requiring some separate assumption to enhance it.
In the oscillatory model, the relative streaming of electrons and positrons allows the 
maximum possible growth rate for the two-stream instability, 
at phases where the counterstreaming is nonrelativistic or mildly relativistic.  
Apart from the two-stream instability, LAE may also operate in conversion
of LAEW to electromagnetic radiation. 

In an oscillatory pulsar magnetosphere, cyclotron resonance can have 
a significant effect on the propagation of the coherent radio emission.
In the conventional polar cap models, for a wave propagating at an angle $\theta$ to the
magnetic field, the cyclotron resonance occurs preferentially in the large-angle regime
$\theta\gg1/\gamma$ at a frequency $\omega=\Omega_e/\gamma\theta^2$, located
at a radius, which is generally in the outer magnetosphere,
$r_c\sim (\Omega_{e0}/\gamma\theta^2\omega)^{1/3}$, where $\Omega_{e0}$ is the 
cyclotron frequency at the surface \citep{lm01}. In the oscillatory pulsar magnetosphere,
the cyclotron resonance can occur at $\omega=\Omega_e/2\gamma$ for particles moving
toward the star. The cyclotron radius $r_c$ varies with oscillating $\gamma$, with 
the smallest radius being $(r_c)_{\rm min}\approx(\Omega_{e0}/2\gamma_{\rm max}\omega)^{1/3}$;
This radius is smaller than in the usual polar cap models by a factor 
$(\theta^2/2)^{1/3}\approx 0.17$ for $\theta=0.1$.

There are some limitations of our model, notably the one-dimensional 
assumption that may not be realistic for an acceleration region extended 
beyond $>R(R/R_{LC})^{1/2}$. The effect of the side wall of the open field line 
region needs to be included in the calculation. Such region can be modeled as a
wave guide and propagation of LAEWs in such wave guide 
will be discussed elsewhere. 
Nonetheless, from this one-dimensional, analytical
model we are able to derive some fundamental features of LAEWs that should remain
valid qualitatively for a more general, three-dimensional case as well.
The fluid treatment adopted here may not be accurate for pulsar plasma as
numerical simulations showed that pairs from a cascade generally have
a broad distribution \citep{ae02} and inclusion of a particle distrbution requires
a kinetic formalism which is beyond the scope of this paper.

\section*{Acknowledgements}
We thank Mike Wheatland for helpful comments.

\newpage

\appendix

\renewcommand{\theequation}{\thesection\arabic{equation}}

\section{Derivation in the temporal gauge}

\setcounter{equation}{0}

In this appendix, we outline an alternative derivation of the wave equation
in the temporal gauge. Electric and magnetic fields can be expressed in terms 
of a vector potential
\begin{equation}
\bE=-{\partial\bA\over\partial t}\,\quad\quad 
\bB=\bnabla\times\bA.
\end{equation}
From (\ref{eq:curlB}) we have
\begin{equation}
\bnabla(\bnabla\cdot\bA)-\nabla^2\bA+{1\over c^2}{\partial^2\bA\over\partial t^2}=\mu_0(\bJ-\bJ_R).
\label{eq:A}
\end{equation}
Assuming that all the relevant quantities are functions of $\chi$, the 
parallel (to $\bkappa$) component of (\ref{eq:A}) takes the form
\begin{equation}
{d^2A_{\parallel}\over d\chi^2}={\mu_0\over\beta^2_V}(J_\parallel-J_{0\parallel}-J_{R\parallel}),
\label{eq:A1}
\end{equation}
with $A_\parallel=\bkappa\cdot\bA$ and
\begin{eqnarray}
J_{0\parallel}&=&\nabla_\parallel(\bnabla\cdot\bA)-\nabla^2A_\parallel\nonumber\\
&=&\nabla_\parallel(\bnabla_\perp\cdot\bA_\perp)-\nabla^2_\perp A_\parallel.
\end{eqnarray}
The equation of motion (\ref{eq:EqMotion}) can be written into a similar form to
(\ref{eq:EqMotion3}):
\begin{equation}
(\beta_V-\bbeta_{s\parallel}){du_{s\parallel}\over d\chi}=
-{se\beta_V\over m_ec}{dA_\parallel\over d\chi}+{1\over c}\left(
{q_{s\parallel}\over m_ec^2}-{Q\over 2N_s}u_{s\parallel}\right),
\end{equation}
which can be integrated to yield
\begin{eqnarray}
{1-\beta_V\beta_s\over(1-\beta^2_s)^{1/2}}-1&=&
{se\beta_V\over m_ec}(A_\parallel-A_{0\parallel})\nonumber\\
&&-{1\over c}\int^\chi_0\left(
{q_s\over m_ec^2}-{Q\over2N_s}u_s\right)\,d\chi',
\label{eq:EqMotion11}
\end{eqnarray}
where $A_{0\parallel}\equiv A_\parallel(\chi=0)$ and $s=\pm$ corresponds to electrons ($+$) and 
positrons ($-$). Assuming $\tilde{A}=(e/m_ec)(A_\parallel-A_{0\parallel})$, for the 
electron component $\beta=\beta_-$, one obtains
\begin{eqnarray}
{\textstyle{1\over2}}\beta^2_V\left({d\tilde{A}\over d\chi}\right)^2
&=&{\omega^2_p\over c^2}
\Biggl[f-{1\over(1-\beta^2)^{1/2}}\nonumber\\
&&\times\biggl(
n_-+g\eta_+-(1-\beta_V\beta){j_{0\parallel}\over\beta_V}\biggr)\Biggr],
  \label{eq:upm4}
\end{eqnarray}
where $f$ is an integration constant.
Clearly, a physical solution requires the RHS to be non-negative; this condition
can be satisfied only if $\beta_{min}\leq \beta\leq\beta_{max}$, where $\beta_{min}$ and
$\beta_{max}$ are the minimum and the maximum velocities at which 
the RHS is zero. The electric field can be found from
\begin{eqnarray}
\tilde{E}_\parallel=-\beta_V{d\tilde{A}\over d\tilde{\chi}}.
\label{eq:E3}
\end{eqnarray}

\section{Electron gas}
\setcounter{equation}{0}

Here we reproduce the known result for a LAEW in an electron gas, by retaining the 
electron component only. Specifically, we set $\beta_0\equiv\beta_{-0}=0$ and
$\eta_+=\eta_0=\eta_{_{GJ}}=0$. 
The maximum velocity can be expressed in terms of $\tilde{E}_0$:
\begin{equation}
{1\over(1-\beta^2_m)^{1/2}}\equiv {\textstyle{1\over2}}\tilde{E}^2_0+1.
\end{equation}
Eq (\ref{eq:EqMotion4}) and (\ref{eq:waveE4}) reproduce
an analytical form similar to that given by \citet{aetal75}: 
\begin{eqnarray}
-\int
\Biggl[{1\over(1-\beta^2_m)^{1/2}}-{1\over(1-\beta^2)^{1/2}}\Biggr]^{-1/2}d\xi
= 2^{1/2}{\omega_p\over c}\chi,
\end{eqnarray}
\begin{equation}
E_\parallel=\pm2^{1/2}{m_ec\over e}\omega_p
\Biggl[
{1\over(1-\beta^2_m)^{1/2}}-
{1\over(1-\beta^2)^{1/2}}
\Biggr]^{1/2}.
\end{equation}
Consider the relativistic limit $\gamma_m\equiv1/(1-\beta^2_m)^{1/2}\gg1$;
Using $\beta\approx \pm(1-\gamma^{-2}/2)$, one finds
\begin{eqnarray}
&&
2\left[1\pm\beta_V\left(1+{1\over\gamma_m}\right)\right](\gamma_m-1)^{1/2}\nonumber\\
&&+
\left[2\mp\beta_V\left(1+{1\over2\gamma\gamma_m}\right)\right](\gamma_m-\gamma)^{1/2}
\nonumber\\
&&\pm{\beta_V\over2\gamma^{3/2}_m}
\left[\arctanh\left(1-{1\over\gamma_m}\right)-
\arctanh\left(1-{\gamma\over\gamma_m}\right)\right]\nonumber\\
&&=\sqrt{2}\left({\omega_p\over c}\right)\chi.
\end{eqnarray}
Since $\beta$ is a periodic function of $\chi$, we can define a period $T$ by
\begin{eqnarray}
&&
2\int^{\beta_m}_{-\beta_m} d\beta
{\beta_V-\beta\over (1-\beta^2)^{3/2}}
\left[{1\over(1-\beta^2_m)^{1/2}}-{1\over(1-\beta^2)^{1/2}}\right]^{-1/2}
\nonumber\\
&&=\sqrt{2}\,\omega_p\beta_VT.
\label{eq:T}
\end{eqnarray} 
The RHS can be written into the form in the $\gamma_m\gg1$ limit:
\begin{equation}
4\int^{\gamma_m}_1{\beta d\gamma\over(\gamma_m-\gamma)^{1/2}}\approx 
8\gamma^{1/2}_m.
\end{equation}
If we define a frequency $\omega=2\pi/T$, (\ref{eq:T}) leads
to
\begin{equation}
\omega={\sqrt{2}\pi\omega_p\over4\gamma^{1/2}_m},
\end{equation} 
which is similar to (\ref{eq:freq}).


\begin{thebibliography}{22}
\bibitem[\protect\citeauthoryear{Akhiezer et al.}{1975}]{aetal75}
Akhiezer, A. I., Akhiezer, I. A., Polovin, R. V., Sitenko, A. G., Stepanov, K. N., 
1975, Plasma Electrodynamics Vol. 2, Pergamon Press
\bibitem[\protect\citeauthoryear{Arendt \& Eilek}{2002}]{ae02}
Arendt, P. N., Eilek, J., 2002, ApJ, 581, 451
\bibitem[\protect\citeauthoryear{Arons \& Scharlemann}{1979}]{as79}
Arons, J., Scharlemann, E., 1979, ApJ, 231, 854
\bibitem[\protect\citeauthoryear{Arons}{1983}]{a83}
Arons, J., 1983, ApJ, 266, 215
\bibitem[\protect\citeauthoryear{Asseo, Kennel \& Pella}{1977}]{aetal77}
Asseo, E., Kennel, C. F., Pella, R., 1977, A\&A, 65, 401
\bibitem[\protect\citeauthoryear{Beloborodov}{2007}]{b07}
Beloborodov, A. M., 2007, astro-ph0710.0920
\bibitem[\protect\citeauthoryear{Blaskiewicz, Cordes \& Wasserman}{1991}]{betal91}
Blaskiewicz, M, Cordes, J., Wasserman, I., 1991, ApJ, 370, 643
\bibitem[\protect\citeauthoryear{Cheng \& Ruderman}{1976}]{cr76}
Cheng, A., Ruderman, M., 1976, ApJ, 203, 209
\bibitem[\protect\citeauthoryear{Cheng, Ho \& Ruderman}{1986}]{cetal86}
Cheng, K. S., Ho, C., Ruderman, M., 1986, ApJ, 300, 500
\bibitem[\protect\citeauthoryear{Dermer}{1990}]{d90}
Dermer, C. D., 1990, ApJ, 360, 197
\bibitem[\protect\citeauthoryear{Erber}{1966}]{e66}
Erber, T., 1966, Rev. Mod. Phys., 38, 626
\bibitem[\protect\citeauthoryear{Everett \& Weisberg}{2001}]{ew01}
Everett, J. E., Weisberg, J. M., 2001, ApJ, 553, 341
\bibitem[\protect\citeauthoryear{Fawley, Arons \& Scharlemann}{1977}]{fetal77}
Fawley, W. M., Arons, J., Scharlemann, E. T., 1977, ApJ, 217, 227
\bibitem[\protect\citeauthoryear{Harding \& Muslimov}{1998}]{hm98}
Harding, A., Muslimov, A., 1998, ApJ, 508, 328
\bibitem[\protect\citeauthoryear{Harding \& Muslimov}{2005}]{hm05}
Harding, A., Muslimov, A., 2005, Ap\&SS, 297, 63
\bibitem[\protect\citeauthoryear{Hirotani}{2006}]{h06}
Hirotani, K., 2006, ApJ, 652, 1475
\bibitem[\protect\citeauthoryear{Levinson et al.}{2005}]{letal05}
Levinson, A., Melrose, D. B., Judge, A., Luo, Q., 2005, ApJ, 631, 456
\bibitem[\protect\citeauthoryear{Luo}{1996}]{l96}
Luo, Q., 1996, ApJ, 468, 338
\bibitem[\protect\citeauthoryear{Luo \& Melrose}{2001}]{lm01}
Luo, Q., Melrose, D. B., 2001, MNRAS, 325, 187
\bibitem[\protect\citeauthoryear{Medin \& Lai}{2007}]{ml07}
Medin Z., Lai D., 2007, Advances in Space Research (in press)
\bibitem[\protect\citeauthoryear{Melrose}{1978}]{m78}
Melrose, D. B., 1978, ApJ, 225, 557
\bibitem[\protect\citeauthoryear{Melrose, Levinson, Judge, \& Luo}{2005}]{metal05}
Melrose, D., Levinson, A., Judge, A., Luo, Q., 2005, AIP Proceedings
\bibitem[\protect\citeauthoryear{Mestel}{1999}]{m99}
Mestel, L., 1999, Stellar Magnetism, Clarendon Press: Oxford
\bibitem[\protect\citeauthoryear{Mestel \& Shibata}{1994}]{ms94}
Mestel, L., Shibata, S., 1994, MNRAS, 271, 621
\bibitem[\protect\citeauthoryear{Michel}{2004}]{m04}
Michel, F. C., 2004, Adv. Space Res. 33, 542
\bibitem[\protect\citeauthoryear{Michel}{1975}]{m75}
Michel, F. C., 1975, ApJ, 197, 193
\bibitem[\protect\citeauthoryear{Muslimov \& Tsygan}{1992}]{mt92}
Muslimov, A. G., Tsygan, A., 1992, MNRAS, 255, 61
\bibitem[\protect\citeauthoryear{Rowe}{1992a}]{r92a}
Rowe, E. T., 1992, Aust. J. Phys. 45, 1
\bibitem[\protect\citeauthoryear{Rowe}{1992b}]{r92b}
Rowe, E. T., 1992, Aust. J. Phys. 45, 21
\bibitem[\protect\citeauthoryear{Romani}{1996}]{r96}
Romani, R., 1996, ApJ, 470, 469 
\bibitem[\protect\citeauthoryear{Ruderman \& Sutherland}{1975}]{rs75}
Ruderman, M. A., \& Sutherland, P. G. 1975, ApJ, 196, 51
\bibitem[\protect\citeauthoryear{Scharlemann \& Wagoner}{1973}]{sw73}
Scharlemann, E. T., Wagoner, R. V., 1973, ApJ, 182, 951
\bibitem[\protect\citeauthoryear{Shibata}{1991}]{s91}
Shibata, S., 1991, ApJ, 378, 239
\bibitem[\protect\citeauthoryear{Shibata}{1997}]{s97}
Shibata, S., 1997, MNRAS, 287, 262
\bibitem[\protect\citeauthoryear{Sturner}{1995}]{s95}
Sturner, S. J., 1995, ApJ, 446, 292
\bibitem[\protect\citeauthoryear{Sturrock}{1971}]{s71}
Sturrock, P. A. 1971, ApJ, 164, 529
\bibitem[\protect\citeauthoryear{Thompson}{2001}]{t01}
Thompson, D. J., 2001, in High Energy Gamma-Ray Astronomy, AIP Proceedings, vol 558, p. 103
\bibitem[\protect\citeauthoryear{Timokhin}{2006}]{t06}
Timokhin, A. N., 2006, MNRAS, 368, 1055
\bibitem[\protect\citeauthoryear{Verdon \& Melrose}{2007}]{vm07}
Verdon, M., Melrose, D. B., 2008, in 40 Years of Pulsars--Millisecond Pulsars, Magnetars,
and More, eds C. G. Bassa, Z. Wang, V. M. Kaspi, AIP Conf. Proc. Vol 983, p. 133
\end{thebibliography}
\end{document}